\documentclass[twocolumn,showpacs,preprintnumbers,amsmath,amssymb]{revtex4-1}

\usepackage{afterpage}
\usepackage[dvipdfmx]{graphicx}
\usepackage{color,bm}

\begin{document}

\title{
Gravitational deflection angle of light: 
Definition by an observer and its application to an asymptotically nonflat spacetime} 
\author{Keita Takizawa}
\author{Toshiaki Ono}
\author{Hideki Asada} 
%\email{asada@phys.hirosaki-u.ac.jp}
\affiliation{
Graduate School of Science and Technology, Hirosaki University,
Aomori 036-8561, Japan} 
\date{\today}

\begin{abstract} 
The gravitational deflection angle of light 
for an observer and source at finite distance from a lens object  
has been studied by Ishihara et al. [Phys. Rev. D, 94, 084015 (2016)], 
based on the Gauss-Bonnet theorem 
with using the optical metric. 
Their approach to finite-distance cases is limited within an asymptotically flat spacetime. 
By making several assumptions, 
we give an interpretation of their definition from the observer's viewpoint: 
The observer assumes the direction of a hypothetical light emission 
at the observer position and 
makes a comparison between the fiducial emission direction and the direction 
along the real light ray. 
The angle between the two directions at the observer location 
can be interpreted as the deflection angle by Ishihara et al. 
The present interpretation does not require the asymptotic flatness. 
Motivated by this, we avoid such asymptotic regions to discuss 
another integral form of the deflection angle of light. 
This form makes it clear that the proposed deflection angle can be 
used not only for asymptotically flat spacetimes 
but also for asymptotically nonflat ones. 
We examine the proposed deflection angle in two models for the latter case; 
Kottler (Schwarzschild-de Sitter) solution in general relativity and 
a spherical solution in Weyl conformal gravity. 
Effects of finite distance on the light deflection in Weyl conformal gravity 
result in an extra term in the deflection angle, 
which may be marginally observable in a certain parameter region. 
On the other hand, those in Kottler spacetime are beyond reach of the current technology. 
\end{abstract}


\maketitle

\section{Introduction}
The gravitational deflection angle of light is of great importance 
in modern gravitational physics, since the pioneering experimental test 
of the gravitational deflection of light was done by Eddington 
\cite{Eddington}. 
The gravitational deflection angle of light plays also a fundamental role 
in the successful observations of gravitational lensing, 
which enable to measure a dark matter distribution,  
extra solar planets and so on. 
Notably, the Event Horizon Telescope team has recently reported 
a direct imaging of the immediate vicinity of 
the central black hole candidate of M87 galaxy 
\cite{EHT}. 

In a conventional derivation of the gravitational deflection angle of light, 
a source (denoted by S) of light and an observer are assumed to be located 
at infinite distance from a lens object (denoted by L). 
Therefore, the asymptotic flatness of the spacetime is required. 
In the rest of this paper, the observer is called the receiver (denoted by R) 
in order to avoid a notational confusion 
between $r_O$ and $r_0$ (the closest approach) by using $r_R$. 
In astronomy, indeed, the distance from the source to the receiver 
is finite. 
Therefore, a lot of attempts have been made to discuss finite-distance effects 
on the gravitational deflection angle of light (e.g. \cite{Sereno}). 

Gibbons and Werner proposed an alternative way of deriving 
the gravitational deflection angle of light 
\cite{GW}. 
They also assumed that the source and receiver are located 
at an asymptotically Minkowskian region. 
They used 
the Gauss-Bonnet theorem \cite{GBMath} 
to a spatial domain described by the optical metric, 
with which a light ray is described as a spatial curve. 
Their method has been largely applied to many spacetime models 
especially by Jusufi and his collaborators \cite{Jusufi2017a,Jusufi2017b,Jusufi2018,Crisnejo2019}. 

In order to investigate finite-distance effects on the light deflection 
in a static and spherically symmetric spacetime, 
Ishihara et al. have proposed a method that extends Gibbons and Werner's idea 
\cite{Ishihara2016}.  
This method by Ishihara et al. has been generalized to study a strong deflection case, 
axisymmetric spacetimes such as the Kerr solution and a rotating wormhole 
and also a deficit angle model 
\cite{Ishihara2017,Ono2017,Ono2018,Ono2019}. 
Their extensions are still limited within asymptotically flat spacetimes, 
because an integration range of the Gaussian curvature includes 
the asymptotically flat region in the space described by the optical metric. 

It is interesting to extend Ishihara et al. method 
to a spacetime that is not asymptotically flat. 
The main results of this paper are two. 
First, by making several assumptions at the receiver location, 
we give an interpretation to their definition of the deflection angle. 
This interpretation does not require the asymptotic flatness. 
Secondly, motivated by this, we avoid such asymptotic regions 
to discuss another integral form of the deflection angle of light 
based on the Gauss-Bonnet theorem. 
Irrespective of whether the spacetime is asymptotically flat, 
the proposed form of the deflection angle 
allows us to describe finite-distance effects 
on the gravitational deflection angle of light. 

This paper is organized as follows. 
In Section II, 
the definition of the deflection angle by Ishihara et al.  
is interpreted at the receiver position. 
We propose another form of the deflection angle based 
on the Gauss-Bonnet theorem, 
such that we can avoid treating asymptotic regions. 
In Section III, 
we examine the proposed definition of the deflection angle of light in two examples; 
Kottler (Schwarzschild-de Sitter) solution in general relativity and 
a spherical solution in Weyl conformal gravity. 
Finite-distance effects in these models are mentioned. 
Section IV is devoted to the conclusion. 
Throughout this paper, we use the unit of $G = c = 1$.

\section{Avoiding the asymptotic region}
\subsection{The deflection angle proposed by Ishihara et al. }
Following Reference \cite{Ishihara2016}, 
we shall briefly summarize the Ishihara et al. method and their definition. 
We consider a static and spherically symmetric spacetime. 
The metric can be written as 
\begin{align}
ds^2 &= g_{\mu\nu} dx^{\mu} dx^{\nu} 
\nonumber\\
&= -A(r) dt^2 + B(r) dr^2 + C(r) d\Omega^2 , 
\label{ds2-SSS-AB}
\end{align}
where 
$d\Omega^2 \equiv d\theta^2 + \sin^2\theta d\phi^2$ and 
$\phi$ is associated with the rotational symmetry.  
This metric form allows for 
a wormhole solution with a throat as well as a black hole spacetime. 
If we choose $C(r) = r^2$, then, $r$ denotes the circumference radius.

\begin{figure}
\includegraphics[width=8.6cm]{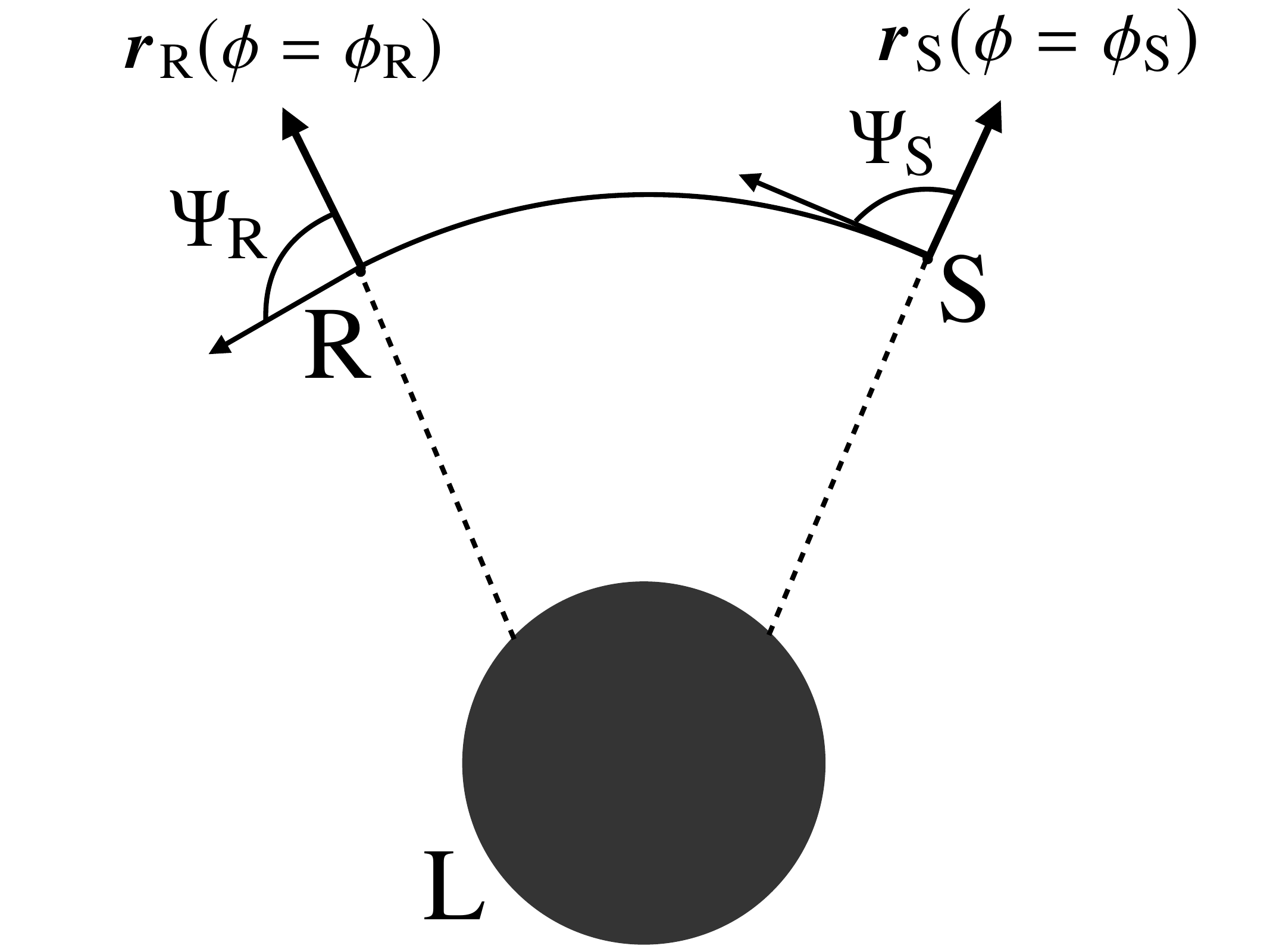}
\caption{
The light ray and radial directions. 
The angle between the light ray and the radial direction at the receiver 
is $\Psi_R$ and 
that at the source is $\Psi_S$. 
The coordinate angle between the receiver and the source 
is $\phi_{RS} = \phi_R - \phi_S$. 
}
\label{fig-configuration}
\end{figure}

Ishihara et al. proposed a definition of the gravitational deflection of light, 
which does not require a receiver and source of light at the infinite distance 
from a lens object \cite{Ishihara2016}. 
See also \cite{Ono2019b} for discussions on 
its observability. 
The definition by Ishihara et al.\cite{Ishihara2016} is 
\begin{align}
\alpha \equiv \Psi_R - \Psi_S + \phi_{RS} , 
\label{alpha}
\end{align}
where $\Psi_S$ and $\Psi_R$ are the angles between the radial direction 
and the light ray at the source position and at the receiver position, respectively, 
and $\phi_{RS}$ is a coordinate angle between the receiver and source. 
See Figure \ref{fig-configuration}. 
For instance,  Rindler and Ishak proposed an alternative definition of the deflection angle 
by their assuming that the lens, receiver and source are aligned 
in the Kottler spacetime \cite{Rindler,Ishak}. 
Their work received a lot of attention \cite{Park,Sereno,Bhadra,Simpson,Ishak2010,AK}. 
Recently, Arakida made an attempt to give another definition of the deflection angle 
\cite{Arakida}. 
However, he compared geodesics which belong to two different spacetimes. 
Therefore, it is an open issue whether his method can be mathematically justified. 
Motivated by Arakida's attempt, Crisnejo, Gallo and Rogers \cite{CGR2019} 
proposed an alternative definition of the deflection angle, 
where the integration domain 
for the Gauss-Bonnet theorem is a finite quadrilateral. 
See also Figure 2 and Eq. (11) in Reference \cite{CGR2019}.  

Without the loss of generality, 
the photon orbital plane can be chosen as the equatorial plane 
($\theta = \pi/2$). 
By using the energy of the photon ($E$) and the angular momentum 
($L$), 
the impact parameter of the light ray is defined as 
\begin{align}
b &\equiv \frac{L}{E} 
\nonumber\\
&= \frac{C(r)}{A(r)} \frac{d\phi}{dt} . 
\label{b}
\end{align}
By using $b$ in the null condition $ds^2 =0$, 
we obtain the orbit equation as 
\begin{align}
\left( \frac{dr}{d\phi} \right)^2 
+ \frac{C(r)}{B(r)} 
&= \frac{\left\{C(r)\right\}^2}{b^2 A(r)B(r)} . 
\label{orbiteq}
\end{align}

$ds^2 = 0$ is solved for the time coordinate as 
\begin{align}
dt^2 &= \gamma_{IJ} dx^I dx^J 
\nonumber\\
&= \frac{B(r)}{A(r)} dr^2 + \frac{C(r)}{A(r)} d\phi^2 ,  
\label{gamma}
\end{align}
where $I$ and $J$ denote $r$ and $\phi$. 
We refer to $\gamma_{IJ}$ as the optical metric. 
$\gamma_{IJ}$ defines a two-dimensional Riemannian space, 
in which the light ray is described as a spatial geodetic curve 
with $\gamma_{IJ}$ but not $g_{IJ}$. 
By using $\gamma_{IJ}$, the angles $\Psi_R$ and $\Psi_S$ in Eq. (\ref{alpha}) are defined. 

\begin{figure}
\includegraphics[width=8.6cm]{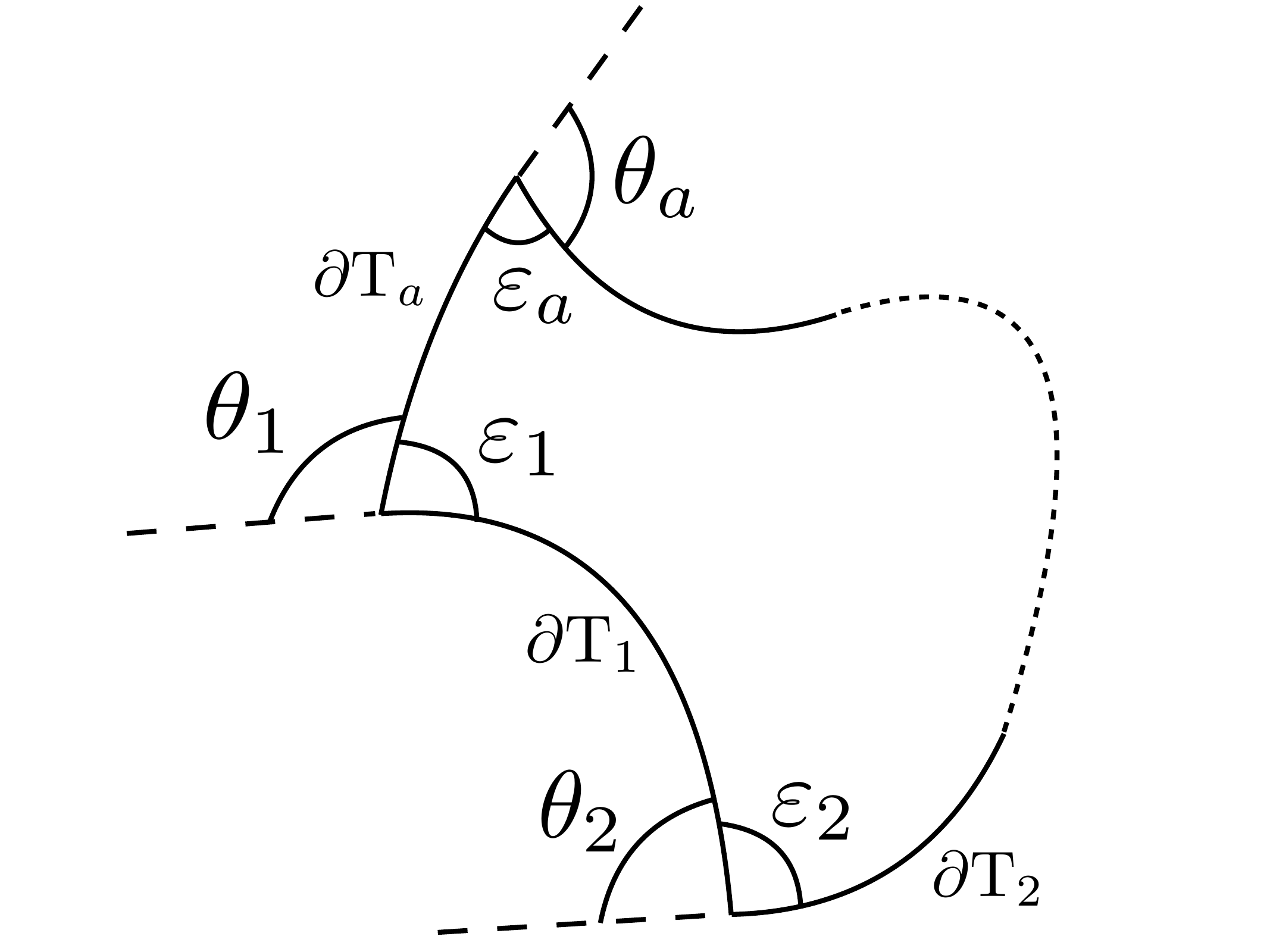}
\caption{
Schematic figure for the Gauss-Bonnet theorem. 
The inner angle is $\varepsilon_a$ and 
the outer (jump) angle is $\theta_a$. 
}
\label{fig-GB}
\end{figure}

The Gauss-Bonnet theorem can be expressed as 
\cite{GBMath}
\begin{align}
\iint_{T} K dS + \sum_{a=1}^N \int_{\partial T_a} \kappa_g d\ell + 
\sum_{a=1}^N \theta_a = 2\pi , 
\label{GB}
\end{align}
where 
$T$ is a two-dimensional orientable surface with boundaries $\partial T_a$ 
($a=1, 2, \cdots, N$) that 
are differentiable curves, 
$\theta_a$ ($a=1, 2, \cdots, N$) denote jump angles, 
$K$ denotes the Gaussian curvature of 
the surface $T$, 
$dS$ is the area element of the surface, 
$\kappa_g$ means the geodesic curvature of $\partial T_a$, 
and $\ell$ is the line element along the boundary. 
See Figure \ref{fig-GB}. 
The sign of the line element is chosen such that it can be 
consistent with the surface orientation.

\begin{figure}
\includegraphics[width=8.6cm]{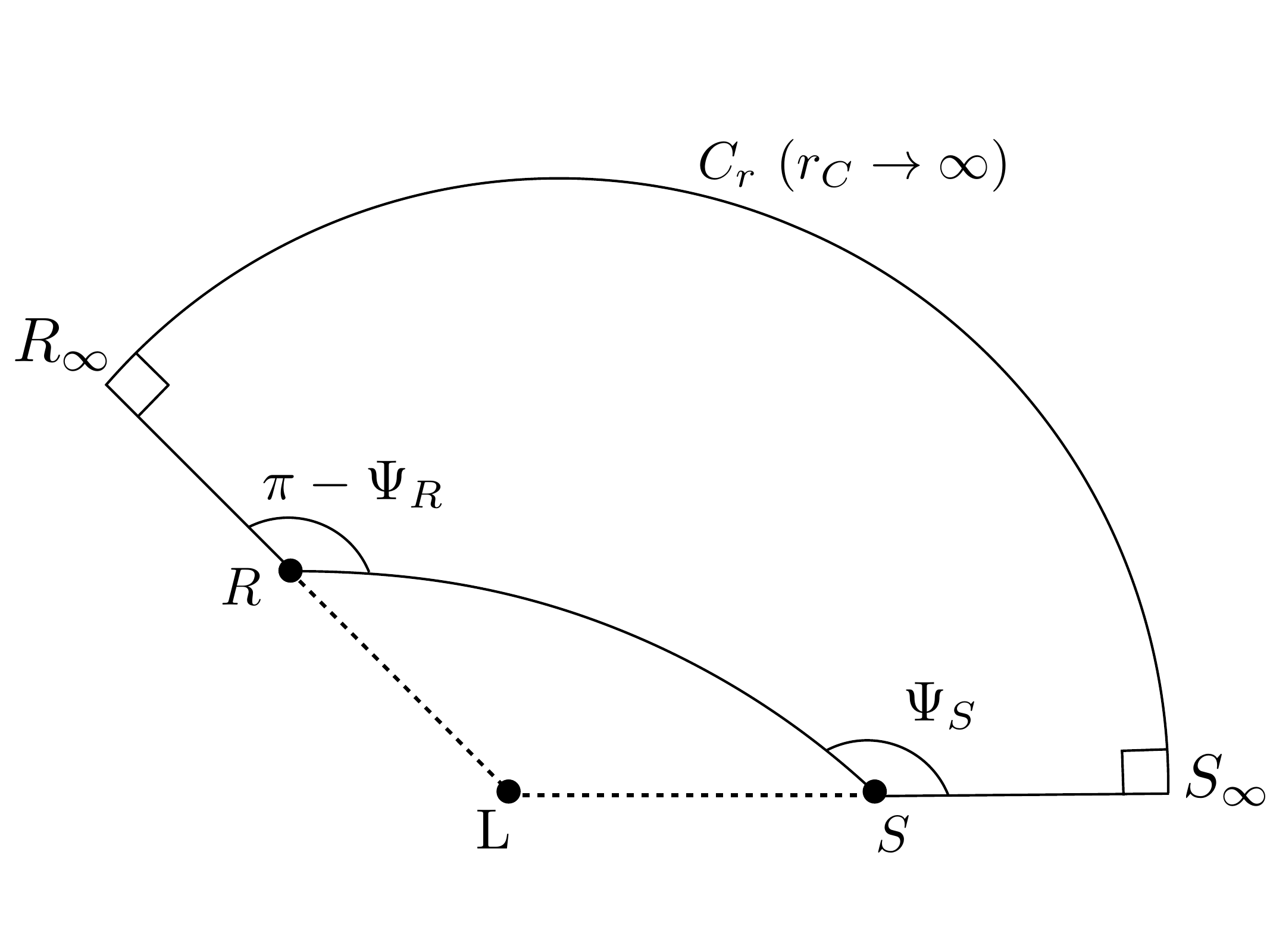}
\caption{
Quadrilateral ${}^{\infty}_{R}\Box^{\infty}_{S}$
embedded in a curved space. 
}
\label{fig-Box}
\end{figure}

Ishihara et al. \cite{Ishihara2016} consider a quadrilateral 
${}^{\infty}_{R}\Box^{\infty}_{S}$, 
which consists of the spatial curve for the light ray, 
two outgoing radial lines from R and from S 
and a circular arc segment $C_r$ 
of coordinate radius $r_C$ ($r_C \to \infty$) 
centered at the lens 
which intersects the radial lines  
through the receiver or the source. 
See Figure \ref{fig-Box}. 
They restrict themselves within the asymptotically flat spacetime;  
$\kappa_g \to 1/r_C$ and $d\ell \to r_C d\phi$ 
as $r_C \to \infty$ (See e.g. \cite{GW}). 
Hence, 
$\int_{C_r} \kappa_g d\ell \to \phi_{RS}$.  
By using these things in the Gauss-Bonnet theorem 
for the domain ${}^{\infty}_{R}\Box^{\infty}_{S}$, 
we obtain 
\begin{align}
\alpha 
&= \Psi_R - \Psi_S + \phi_{RS} 
\nonumber\\
&= - \iint_{{}^{\infty}_{R}\Box^{\infty}_{S}} K dS . 
\label{alpha-K}
\end{align}
Eq. (\ref{alpha-K}) shows that $\alpha$ 
is invariant in differential geometry 
and $\alpha$ is well-defined even if L is a singularity point, 
because the domain ${}^{\infty}_{R}\Box^{\infty}_{S}$ 
does not contain the point L. 
It follows that $\alpha=0$ in Euclidean space. 
This is because $K$ vanishes in Euclidean space and the area integral of $K$ 
thus vanishes.

\subsection{Interpretation of the deflection angle of light by Ishihara et al.}
Is Eq. (\ref{alpha}) the deflection angle of light? 
Yes,  but Ishihara et al. have not clearly discussed this issue. 
Therefore, we shall discuss it here. 

Rigorously speaking, the parallel transport of the photon direction is done 
by following the null geodesic. 
Consequently, there is no deflection of light in the full description of the null geodesic. 
When we discuss the deflection of light, therefore, 
we have to introduce a reference direction and 
compare the difference between the fiducial direction and 
the real light ray. 
This procedure imitates the Eddington experiment which 
made, at the receiver position, 
a comparison between the observed (lensed) image direction 
and the intrinsic (unlensed) source direction. 

There can be various and different definitions of such a reference direction 
in a general curved space. 
In this paper, we define a reference direction in the following manner. 
Where is a suitable spatial position for defining the deflection angle? 
From the viewpoint of observations, it would be better to 
choose the receiver position rather than the source position. 
Therefore, we allow the receiver to define the reference direction. 

The angles $\Psi_R$ and $\Psi_S$ have been already defined. 
In principle, the receiver can measure $\Psi_R$ 
and the source can measure $\Psi_S$. 
Is $\Psi_S$ measured by the receiver? 
No for an isotropic emission of light. 
However, yes for an anisotropic radiation from the source in principle, 
as two of the present authors (Ono and Asada) argued \cite{Ono2019b}. 
Hence, we imagine that the receiver knows the direction angle $\Psi_S$ 
of the light emission. 

\begin{figure}
\includegraphics[width=8.6cm]{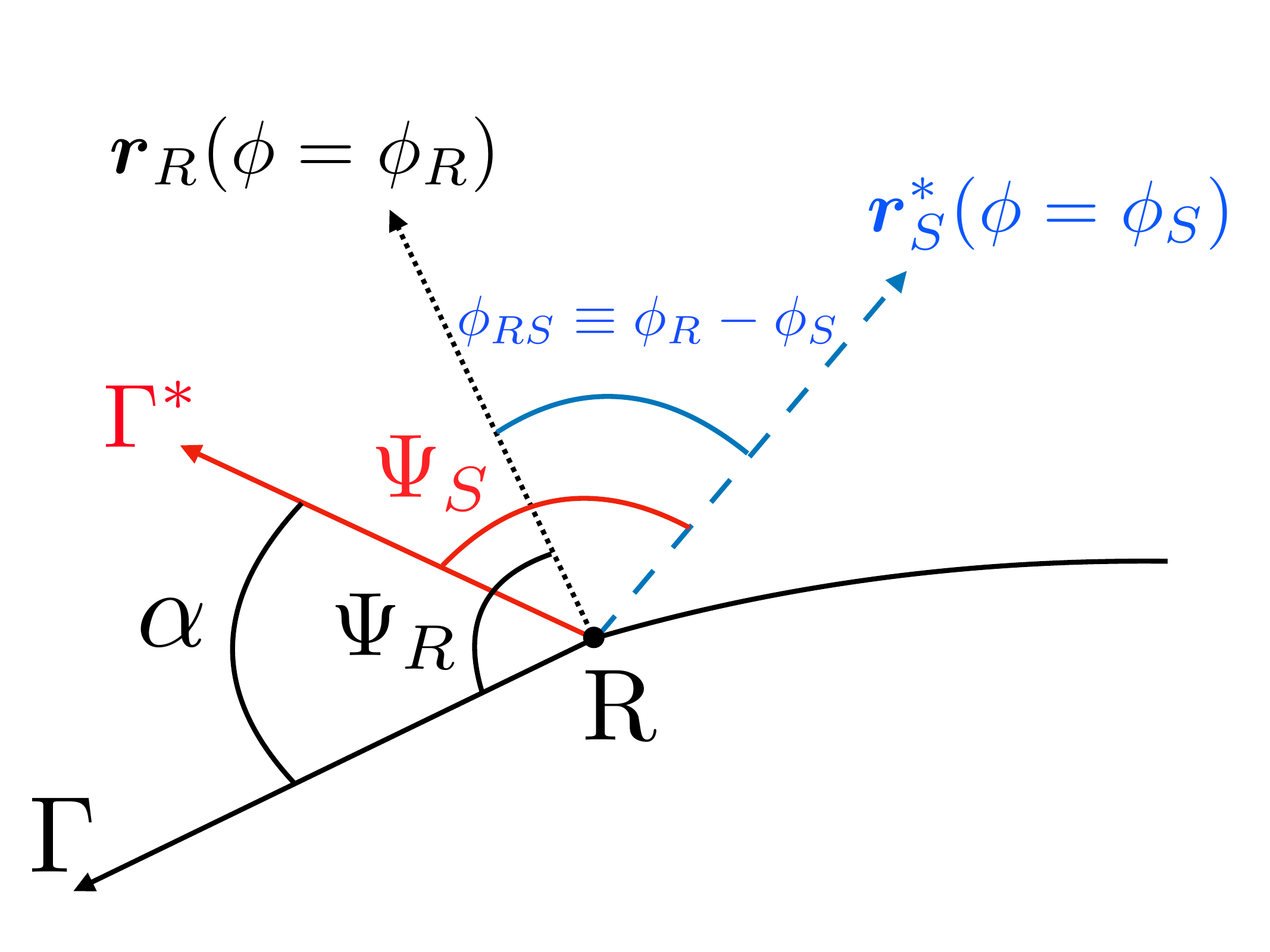}
\caption{
Angles and directions at the receiver. 
The (black in color) dotted line corresponds to the radial direction 
$\bm{r}_R$ at the receiver. 
The (black in color) solid line denotes the tangent vector $\Gamma$ 
to the light ray through the receiver. 
The (blue in color) dashed line means the fiducial radial direction $\bm{r}_S^{\ast}$ 
that is assumed by the receiver. 
The (red in color) solid line is the hypothetical emission direction $\Gamma^{\ast}$ 
that is defined by the receiver. 
The angle between the radial direction $\bm{r}_R$ and the light ray $\Gamma$ 
is denoted as $\Psi_R$. 
The receiver defines the fiducial radial direction $\bm{r}_S^{\ast}$, 
such that 
the angle between $\bm{r}_R$ and $\bm{r}_S^{\ast}$ 
can be the same as $\phi_{RS}$. 
The receiver defines also the hypothetical emission direction $\Gamma^{\ast}$, 
such that the angle between $\Gamma^{\ast}$ and $\bm{r}_S^{\ast}$ 
can be the same as $\Psi_S$. 
As a result, the angle between the light ray $\Gamma$ 
and the hypothetical emission direction $\Gamma^{\ast}$ 
is $\Psi_R - \Psi_S + \phi_{RS}$. 
}
\label{fig-emission}
\end{figure}

Let us define a reference direction at the receiver point. 
See Figure \ref{fig-emission}. 
First, we assume also that the receiver knows the relative angular position $\phi_{RS}$ 
between the source and receiver in the spacetime, 
for instance from very precise ephemeris. 
At the receiver, the radial direction from the lens object 
is denoted by $\bm{r}_R$. 
The radial direction at the source is denoted by $\bm{r}_S$. 
Instead of $\bm{r}_S$, the receiver defines a fiducial direction $\bm{r}_S^{\ast}$  
by rotating the vector $\bm{r}_R$ with the angle $\phi_{RS}$. 

If the space is Euclid, $\bm{r}_S^{\ast}$ equals to $\bm{r}_S$. 
This can be easily understood, when we consider, in a Euclidean space, 
a triangle with vertices at the lens, receiver and source 
and we use the parallel transport of a line. 
In a curved space, the receiver may use the reference direction $\bm{r}_S^{\ast}$ 
as if it were corresponding to $\bm{r}_S$, 
though $\bm{r}_S^{\ast}$ has no direct relation with $\bm{r}_S$ in a curved space. 

We denote the direction of the true light ray at the receiver  
by $\Gamma$. 
Let the receiver define  
a fiducial direction (denoted as $\Gamma^{\ast}$) of the light emission 
at the receiver position. 
$\Gamma^{\ast}$ is defined by assuming  
that the angle between the reference radial direction $\bm{r}_S^{\ast}$ 
and the fiducial emission direction $\Gamma^{\ast}$ 
equals to $\Psi_S$.

The receiver can say that the deflection angle of light $\alpha$ 
is the angle between the light propagation direction $\Gamma$ 
and the fiducial emission direction $\Gamma^{\ast}$. 
Note that $\Gamma$ and $\Gamma^{\ast}$ 
are defined at the receiver position. 
By construction, $\alpha$ becomes $\Psi_R - \Psi_S + \phi_{RS}$. 
This is an interpretation of $\alpha$.

\subsection{Another integral form of the deflection angle of light}
\begin{figure}
\includegraphics[width=8.6cm]{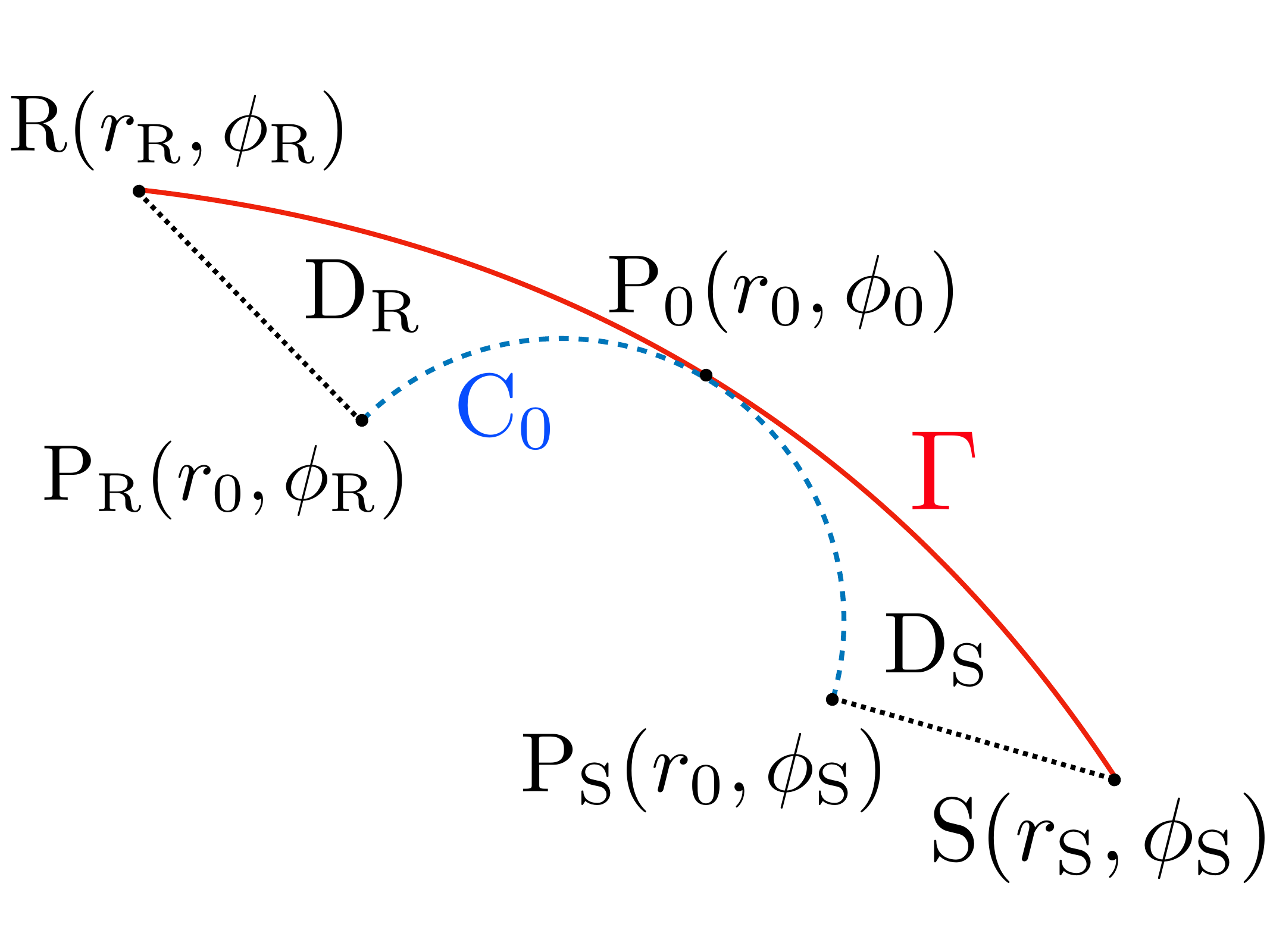}
\caption{
$D_R$ and $D_S$. 
$D_R$ is a trilateral specified by the points $R$, $P_0$ and $P_R$. 
$D_S$ is that specified by the points $S$, $P_0$ and $P_S$. 
}
\label{fig-newdomain}
\end{figure}

The above interpretation of the deflection angle does not require 
the asymptotic flatness, while the integral in Eq. (\ref{alpha-K}) 
needs $r \to \infty$. 
This motivates us to reexamine Eq. (\ref{alpha-K}). 
We consider the following region that is defined 
by using the receiver, source and the closest approach 
of the light ray. 
See Figure \ref{fig-newdomain} for this domain. 
The boundaries of this region are 
the radial lines through $R$ or $S$, 
the light ray from $S$ to $R$, 
and the circle arc segment   
$C_0$ 
with radius of the closest approach $r_0$. 
This domain is divided into two trilaterals. 
We denote one trilateral containing the receiver point by $D_R$ 
and the other trilateral containing the source point by $D_S$. 
In both of them, the inner angle at the closest approach is zero, 
because the radial coordinate $r$ in a light ray 
has a minimum at the closest approach.

For the domain $D_R$ in Figure \ref{fig-newdomain}, 
we use the Gauss-Bonnet theorem to obtain 
\begin{align}
\iint _{D_R} K dS + \int_{P_R}^{P_0} \kappa_g d\ell - \Psi_R + \frac{\pi}{2} = 0 , 
\label{DR}
\end{align}
where $P_0$ denotes the periastron (the closest approach). 
In the similar manner for $D_S$, we obtain 
\begin{align}
\iint _{D_S} K dS + \int_{P_0}^{P_S} \kappa_g d\ell + \Psi_S - \frac{\pi}{2} = 0 . 
\label{DS}
\end{align}

By combining Eqs. (\ref{DR}) and (\ref{DS}), we find 
\begin{align}
\Psi_R - \Psi_S 
= 
\iint _{D_R + D_S} K dS + \int_{P_R}^{P_S} \kappa_g d\ell . 
\label{DR+DS}
\end{align}

Hence, $\alpha$ in Eq. (\ref{alpha}) is rewritten as 
\begin{align}
\alpha
= 
\iint _{D_R + D_S} K dS + \int_{P_R}^{P_S} \kappa_g d\ell + \phi_{RS} . 
\label{alpha-new}
\end{align}
The right-hand side of this equation  
contains the radial coordinate $r \in [r_0, r_R]$ or $[r_0, r_S]$. 
Indeed, this radial interval is exactly the same as that for 
the light ray from the source to the receiver. 
On the other hand, the integration range for the radial coordinate 
in Ishihara et al. is $r \in [r_0, \infty]$ . 
Therefore, they needed the asymptotic flatness 
for treating $r \sim \infty$. 
The new form of $\alpha$ by Eq. (\ref{alpha-new}) 
is better than the previous form by Eq. (\ref{alpha-K}), 
in the sense that Eq. (\ref{alpha-new}) does not require the asymptotic flatness. 
Note that neither Eq. (\ref{alpha-K}) nor Eq. (\ref{alpha-new}) 
needs any constraint on the lens point L. 
Namely, the lens object in the present method can be a black hole with a horizon 
or a wormhole with a throat. 

Please see the next section for Eq. (\ref{alpha-new}) 
in asymptotically nonflat spacetime models.

\section{Examples in asymptotically nonflat spacetimes}
\subsection{Kottler spacetime}
As a first example of asymptotically nonflat spacetimes, 
we consider the Kottler solution \cite{Kottler}. 
The line element is 
\begin{align}
  d s^2 =& -\left(1 - \frac{r_g}{r} - \frac{\Lambda}{3}r^2\right) d t^2
 + \left(1 - \frac{r_g}{r} - \frac{\Lambda}{3}r^2\right)^{-1} d r^2
 \nonumber\\
& + r^2( d\theta^2 + \sin^2 \theta d\phi^2) , 
\label{eq-Kottler}
\end{align}
where $r_g$ and $\Lambda$ are constants. 
For this spacetime, the optical metric on the equatorial plane ($\theta = \pi/2$) 
becomes 
\begin{align}
  \gamma_{rr} &= \left(1 - \frac{r_g}{r} - \frac{\Lambda}{3}r^2\right)^{-2} , 
  \\
  \gamma_{\phi \phi} &= r^2 \left(1 - \frac{r_g}{r} - \frac{\Lambda}{3}r^2\right)^{-1} . 
\end{align}
In the following, we use the linear approximations in $r_g$ and $\Lambda$. 
We thus neglect $O(r_g^2)$ and $O(\Lambda^2)$. 

The Gaussian curvature is calculated as 
\begin{align}
  K &= \frac{R_{r \phi r \phi}}{\det(\gamma_{IJ})}
  \nonumber\\
  &=  -\frac{r_g}{r^3} + \frac{r_g\Lambda}{r} - \frac{\Lambda}{3} 
  + O\left(r^2_g, \Lambda^2 \right) , 
 \label{K-Kottler}
\end{align}
where the Riemann tensor is calculated by using $\gamma_{IJ}$. 
The area element on the equatorial plane is 
\begin{align}
  d S &= \sqrt{\det (\gamma_{IJ})} dr d\phi 
  \notag\\
  &= r\left(1 - \frac{r_g}{r} - \frac{\Lambda}{3}r^2\right)^{-\frac{3}{2}} dr d\phi . 
\label{dS-Kottler}
\end{align}

We define $u$ as the inverse of $r$. 
The photon orbit equation is written as 
\begin{align}
  \left(\frac{d u}{d\phi}\right)^2 = \frac{1}{b^2} - u^2 + r_g u^3 + \frac{\Lambda}{3} . 
  \label{OE-Kottler}
  \end{align}
  The iterative solution is obtained as 
  \begin{align}
   u =& \frac{\sin\phi}{b} + \frac{r_g}{2b^2} (1 + \cos^2 \phi) 
   \notag\\
   &+ \frac{b\Lambda}{6}\sin\phi + \frac{r_g\Lambda}{6}(1 + \cos^2 \phi) 
   + O(r^2_g, \Lambda^2) . 
\label{u-Kottler}
 \end{align}

By using the above equations, 
we obtain 
\begin{align}
  \int_{D_R + D_S} K dS 
  =&
  \frac{r_g}{b}\left(\sqrt{1 - b^2 u^2_S} + \sqrt{1 - b^2 u^2_R}\right) 
\notag\\
  &- \frac{b\Lambda}{6}\left(\frac{\sqrt{1 - b^2 u^2_S}}{u_S}
  + \frac{\sqrt{1 - b^2 u^2_R}}{u_R}\right) \notag\\
  &+ \frac{r_gb\Lambda}{12}\left(\frac{1}{\sqrt{1 - b^2 u^2_S}}
  + \frac{1}{\sqrt{1 - b^2 u^2_R}}\right)
\notag\\
  &+ \left(\frac{b^2 \Lambda}{6} - \frac{r_g}{b} - \frac{r_g}{3}b\Lambda\right)\phi_{RS}
\notag\\
  &  + O(r^2_g, \Lambda^2) . 
\label{intK-Kottler}
\end{align}

The geodesic curvature along a line is \cite{Ono2017}
\begin{align}
  \kappa_g &= \epsilon_{ijk} N^i A^j T^k , 
  \label{kappa}
\end{align} 
where 
$\epsilon_{ijk}$ is the Levi-Civita tensor, 
$N_i$ denotes the unit normal vector to the surface, 
$A^j$ is the acceleration vector of the line, 
and $T^k$ means the unit tangent vector along the line. 
For the circle arc segment with radius $r_0$, 
$A^j$ is the acceleration vector of the arc segment, 
and $T^k$ is the unit tangent vector along the arc. 
The exact form of the geodesic curvature becomes   
\begin{align}
\kappa_g
    &= \frac{1}{r^2_0}\left(r_0 - \frac{3}{2}r_g\right)  .
\end{align}

\begin{align}
    \int^{P_S}_{P_R} \kappa_g d\ell 
    =& -\phi_{RS}
    + \frac{r_g}{b}\phi_{RS} + \frac{r_gb\Lambda}{3}\phi_{RS}
    - \frac{b^2 \Lambda}{6} \phi_{RS} 
    \notag\\
    &+ O\left(r^2_g, \Lambda^2\right) . 
\label{intkappa-Kottler}
\end{align}

Combining Eqs. (\ref{intK-Kottler}) and (\ref{intkappa-Kottler}) 
leads to the deflection angle as 
\begin{align} 
\alpha_{Kottler} 
=&
  \frac{r_g}{b}\left(\sqrt{1 - b^2 u^2_S} + \sqrt{1 - b^2 u^2_R}\right)
\notag\\
& - \frac{b\Lambda}{6}\left(\frac{\sqrt{1 - b^2 u^2_S}}{u_S}
  + \frac{\sqrt{1 - b^2 u^2_R}}{u_R}\right) 
\notag\\
& + \frac{r_g b\Lambda}{12}\left(\frac{1}{\sqrt{1 - b^2 u^2_S}}
  + \frac{1}{\sqrt{1 - b^2 u^2_R}}\right)
\notag\\
&  + O\left(r^2_g, \Lambda^2\right) , 
\label{alpha-Kottler}
\end{align}
where the leading order terms in $\phi_{RS}$ cancel out. 
Eq. (\ref{alpha-Kottler}) fully agrees with Eq. (37) in Ishihara et al. (2016) 
\cite{Ishihara2016}. 
Note that Ishihara et al. (2016) obtained this equation 
by using the form of $\Psi_R - \Psi_S + \phi_{RS}$ but 
not by using their integral form as 
$\int_{{}^{\infty}_{R}\Box^{\infty}_{S}} K dS$. 
Note that their integral diverges. 
This divergent behavior is merely because the integral  in Ishihara (2016) 
is ill-defined for the asymptotically nonflat cases. 
In Kottler spacetime, $K$ diverges as $r \to \infty$.

We briefly discuss how the finite distance affects the light deflection in Kottler spacetime. 
For the simplicity, we assume $u_R = u_S \equiv U$ and 
focus on the coupling between the mass and the cosmological constant. 
A function $V(U)$ for this coupling is defined 
as the third line of Eq. (\ref{alpha-Kottler}) 
\begin{align}
V(U) \equiv \frac{r_g b\Lambda}{6 \sqrt{1 - b^2 U^2}} . 
\label{V}
\end{align}

\begin{figure}
\includegraphics[width=8.6cm]{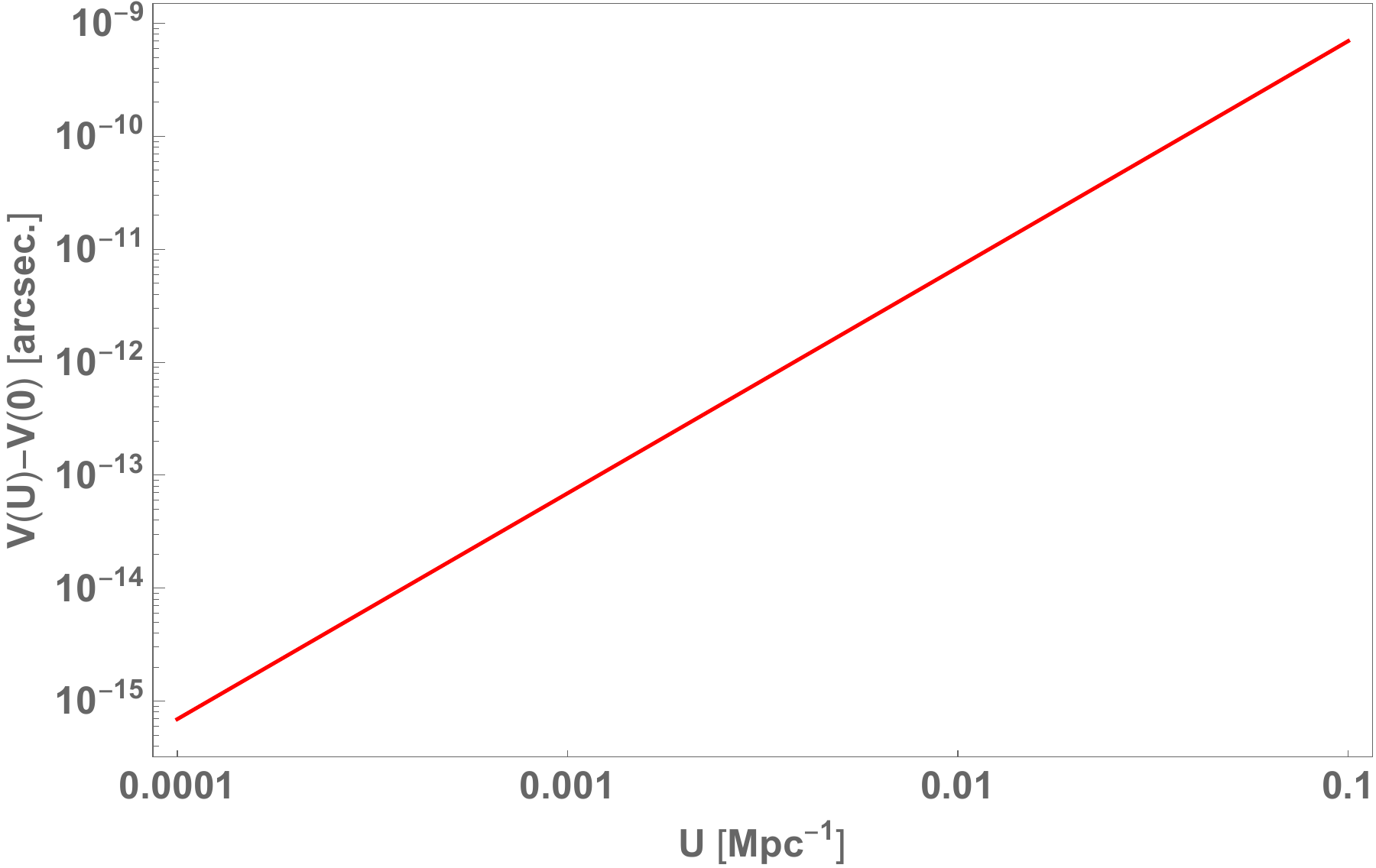}
\caption{
Effects of the finite distance on the deflection of light in Kottler spacetime. 
The vertical axis denotes the difference $V(U) - V(0)$, 
where $u_R = u_S = U$ and $V(U)$ is defined by Eq. (\ref{V}). 
The horizontal axis is $U$, for which we consider a range as 
$\sqrt{\Lambda} < U < (10b)^{-1}$. 
Here, we assume that 
the lens is a cluster of galaxies with mass $10^{15} M_{\odot}$, 
$\Lambda = (10 \mbox{Gpc})^{-2}$, 
and $b = 1$ Mpc for the simplicity. 
The values of $U$ to 0.1 - 1000 $\mbox{Mpc}{}^{-1}$ mean that the receiver and source
are located within 10Mpc - 1kpc from the lens. This situation is not realistic in astronomy. Therefore, we do not consider these values of $U$ in this plot. 
}
\label{fig-difference-Kottler}
\end{figure}

Please see Figure \ref{fig-difference-Kottler} for the difference as $V(U) - V(0)$. 
Note that Kottler solution does not allow $r \to \infty$ ($U \to 0$) but 
$V(U)$ is finite as $r \to \infty$. 
This figure shows how the light deflection at finite-distance receiver and source 
differs from that when the receiver and source are at infinity. 
In this figure, we assume that the lens mass is $m \sim 10^{15} M_{\odot}$ 
(as a cluster of galaxies), 
$\Lambda = (r_H)^{-2} \sim 10^{-52} \mbox{m}^{-2}$, 
and $b = 1$ Mpc, 
where $r_H$ denotes the Hubble horizon radius 
$\sim 10$ Gpc. 
The largest difference of $V(U) - V(0)$ is $\sim 10^{-3}$ microarcseconds  
for $r_R = r_S \sim b$. 
Even the largest difference is thus beyond reach of the current technology. 
Here, we ignore the pure de-Sitter term (the second line) in Eq. (\ref{alpha-Kottler}) 
for the deflection angle in the Kottler model. 
This term does not make any sense at infinity ($r_R, r_S \to \infty$) 
and indeed it diverges there.  
However, we should note that the coupling part between $r_g$ and $\Lambda$ 
(the third line) in Eq. (\ref{alpha-Kottler}) remains finite at infinity, 
though it may make no physical meaning there. 
For readers' convenience, 
we rearrange only the coupling part in Eq. (\ref{alpha-Kottler}) as 
\begin{align}
&\frac{r_g b\Lambda}{12}\left(\frac{1}{\sqrt{1 - b^2 u^2_S}}
  + \frac{1}{\sqrt{1 - b^2 u^2_R}}\right) 
\notag\\
&
= \frac{r_g b\Lambda}{6} 
+ 
\frac{r_g b\Lambda}{12}\left(\frac{1}{\sqrt{1 - b^2 u^2_S}}
  + \frac{1}{\sqrt{1 - b^2 u^2_R}} -2\right) , 
\end{align}
where the first term in the second line is the deflection at infinity  
and the second term denotes the finite-distance effect.

We mention also M87$^*$, the central black hole candidate of M87. 
The finite $r_S$ is fully taken into account in general relativistic simulations 
around the black hole. 
Therefore, we consider finite distance to the receiver in this paragraph. 
We focus on the coupling term containing $r_R$, $r_g$ and $\Lambda$, 
which is expressed by the second term in the third line of Eq. (\ref{alpha-Kottler}).  
This term is 
$\sim 10^{-26} $
where we assume 
$r_R \sim 16 \mbox{Mpc} \sim 10^{23} \mbox{m}$,  
$r_g \sim 7 \times 10^9 M_{\odot} \sim 10^{13} \mbox{m}$, 
$b \sim 10 r_g \sim 10^{14} \mbox{m}$, 
$\Lambda \sim 10^{-52} \mbox{m}^{-2}$.  
Therefore, the effect by finite $r_R$ in Kottler model can be safely ignored 
also for M87$^*$.

\subsection{Spherical solution in Weyl gravity}
For the next example,  
we consider a static and spherically symmetric solution 
in the conformal Weyl gravity \cite{MK}. 
The metric for this solution is 
\begin{align}
d s^2 = -B(r) d t^2 + B^{-1}(r) d r^2 + r^2 (d\theta^2 + \sin^2 \theta d\phi^2) . 
\label{ds-Weyl}
\end{align}
Here, we assume $m \ll 1/\gamma$. 
Namely, $m \gamma \ll 1$. 
Therefore, we can neglect $m^2 \gamma (= m \times m\gamma \ll m)$. 
$B(r)$ can be approximated as 
\begin{align}
B(r) = 1 - 3m \gamma - \frac{2m}{r} + \gamma r -k r^2 . 
\end{align}
For the conformal Weyl gravity, Birkoff's theorem was proven 
\cite{Riegert}. 
Several authors made attempts to calculate the deflection angle in this spacetime 
\cite{Edery,Pireaux2004a,Pireaux2004b,Sultana,Cattani,Ishihara2016}. 

In the Kottler spacetime, we have studied the squared-$r$ term in the metric. 
For simplifying our analysis, we neglect $k$ to choose $k = 0$.

For this spacetime, the optical metric on the equatorial plane ($\theta = \pi/2$) 
becomes 
\begin{align}
  \gamma_{rr} &= [B(r)]^{-2} , 
  \\
  \gamma_{\phi \phi} &= r^2 [B(r)]^{-1} . 
\end{align}

In the following, 
we consider the small deflection and weak field approximations. 
Namely, $m \ll b \ll r_R, r_S$ and $r_R, r_S \ll 1/\gamma$. 
We focus on the linear-order effects by $\gamma$. 
Hence, we neglect $O(\gamma^2)$.

The Gaussian curvature is calculated as 
\begin{align}
  K =& -\frac{2m}{r^3}  - \frac{3m \gamma}{r^2} 
+ O\left( m^2, \gamma^2 \right) , 
 \label{K-Weyl}
\end{align}
where the Riemann tensor is calculated by using $\gamma_{IJ}$. 
The area element on the equatorial plane is 
\begin{align}
  d S &= \sqrt{\det (\gamma_{IJ})} dr d\phi 
  \notag\\
  &= r [B(r)]^{-\frac{3}{2}} d r d\phi . 
\label{dS-Weyl}
\end{align}

The photon orbit equation becomes 
\begin{align}
  \left(\frac{d u}{d\phi}\right)^2 = 
  \frac{1}{b^2} - u^2 + 2mu^3 +3m\gamma u^2 - \gamma u . 
  \label{OE-Weyl}
  \end{align}
  The iterative solution is obtained as 
  \begin{align}
   u =& \frac{\sin\phi}{b} + \frac{m}{b^2}(1 + \cos^2 \phi) - \frac{\gamma}{2} 
 + O(m^2 , \gamma^2) , 
\label{u-Weyl}
 \end{align}
 where terms at $O(m\gamma)$ do not appear. 

By using the above equations, 
we obtain 
\begin{align}
  \int_{D_R + D_S} K dS 
  =& 
  \frac{2m}{b}\left(\sqrt{1 - b^2 u^2_S} + \sqrt{1 - b^2 u^2_R}\right) 
\notag\\
   &   - m\gamma\left(\frac{bu_S}{\sqrt{1 - b^2 u^2_S}} 
   + \frac{bu_R}{\sqrt{1 - b^2 u^2_R}}\right) 
\notag\\
   &   - \frac{2m}{b}\phi_{RS} 
\notag\\
   & + O(m^2 , \gamma^2) . 
\label{intK-Weyl}
\end{align}

The exact form of the geodesic curvature is calculated as  
\begin{align}
\kappa_g
=& 
\frac{1}{r_0}\left(1 - \frac{3m}{r_0} - 3m\gamma + \frac{\gamma}{2}r_0\right)  .
\end{align}
The integral of the geodesic curvature is 
\begin{align}
    \int^{P_S}_{P_R} \kappa_g d\ell 
    =& -\phi_{RS} + \frac{2m}{b}\phi_{RS}
+ O\left(m^2, \gamma^2\right) .
\label{intkappa-Weyl}
\end{align}

Combining Eqs. (\ref{intK-Weyl}) and (\ref{intkappa-Weyl}) 
leads to the deflection angle as 
\begin{align} 
\alpha_{Weyl} 
=&
  \frac{2m}{b}\left(\sqrt{1 - b^2 u^2_S} + \sqrt{1 - b^2 u^2_R}\right)
\notag\\
&   - m\gamma\left(\frac{bu_S}{\sqrt{1 - b^2 u^2_S}} 
   + \frac{bu_R}{\sqrt{1 - b^2 u^2_R}}\right)
  \notag\\
&  + O\left(m^2, \gamma^2\right) , 
\label{alpha-Weyl}
\end{align}
where terms at $O(\gamma)$ do not appear. 
Eq. (\ref{alpha-Weyl}) coincides with Eq. (42) in Ishihara et al. (2016) 
\cite{Ishihara2016}. 
Note that Ishihara et al. (2016) obtained the same equation  
through $\Psi_R - \Psi_S + \phi_{RS}$ but 
not the integral form as 
$\int_{{}^{\infty}_{R}\Box^{\infty}_{S}} K dS$. 
Note that this integral diverges. 
This divergent behavior is merely because the integral domain in Ishihara (2016) 
is ill-defined for the asymptotically nonflat cases. 
For the spherical model in Weyl gravity, 
$K$ remains finite but its area integral diverges as $r \to \infty$ 
\cite{Ishihara2016}. 

Note that the $m\gamma$ coupling term in Eq. (\ref{alpha-Weyl}) 
vanishes in the limit as $r_R, r_S \to \infty$. 
Therefore, this coupling term does not appear in the literature 
\cite{Pireaux2004a,Pireaux2004b,Sultana,Cattani} before 
Reference \cite{Ishihara2016}. 
Ishihara et al. (2016) \cite{Ishihara2016} suggests the existence of this term. 
However,  their integral diverges and hence it has been an open issue 
whether this term exists. 
By using the well-defined integral, the present paper confirms that 
the $m\gamma$-coupling term results from the finite distance.

\begin{figure}
\includegraphics[width=8.6cm]{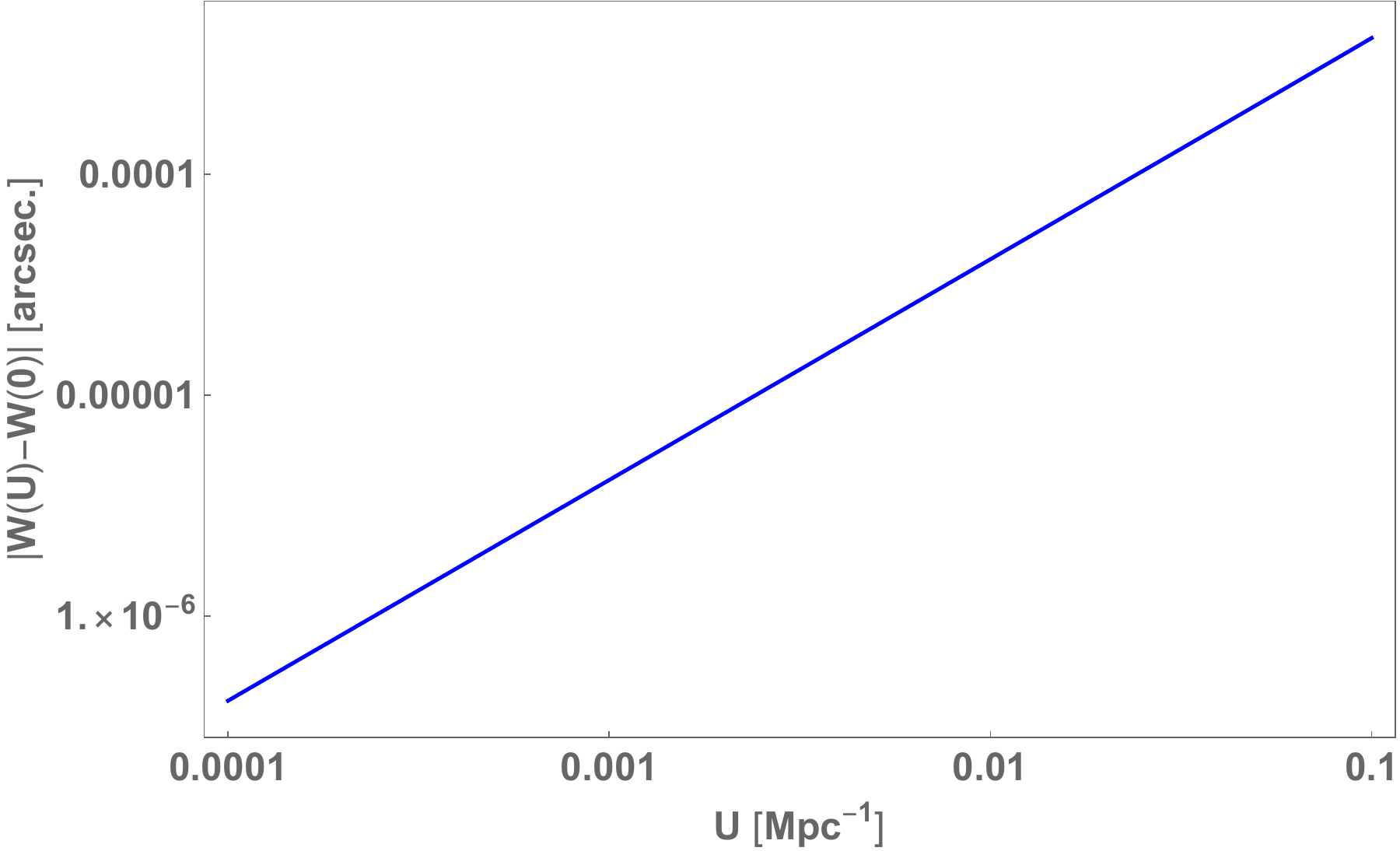}
\caption{
Effects of the finite distance on the deflection of light in Weyl gravity. 
The vertical axis denotes the difference $|W(U) - W(0)|$, 
where $u_R = u_S = U$ and $W(U)$ is defined by Eq. (\ref{W}). 
The horizontal axis is $U$, for which we consider a range as 
$\gamma < U < (10b)^{-1}$. 
Here, we assume that the lens mass is $10^{15} M_{\odot}$, 
$\gamma = (10 \mbox{Gpc})^{-1}$, 
and $b = 1$ Mpc for the simplicity. 
}
\label{fig-difference-Weyl}
\end{figure}

Finally, we discuss how the finite distance affects the light deflection 
for the spherical model in Weyl gravity. 
We assume $u_R = u_S \equiv U$ and 
focus on the coupling 
between the mass and the $\gamma$ parameter. 
A function $W(U)$ due to this coupling 
is defined by the second line of Eq. (\ref{alpha-Weyl}) as 
\begin{align}
W(U) \equiv  - 2 m\gamma \frac{b U}{\sqrt{1 - b^2 U^2}}  .
 \label{W}
\end{align}
$W(U)$ is negative if $\gamma >0$. 
This means that this term is a negative correction to the deflection angle 
if $\gamma$ is positive. 

Please see Figure \ref{fig-difference-Weyl} for the difference as $|W(U) - W(0)|$, 
where $W(0) = 0$.  
This figure shows how the light deflection at finite-distance receiver and source 
differs from that when the receiver and source are at infinity. 
In this figure, we assume that the lens mass is $10^{15} M_{\odot}$, 
$\gamma= (r_H)^{-1}$, 
and $b = 1$ Mpc. 
The largest difference of $|W(U) - W(0)|$ is 
$\sim O(10^2)$ microarcseconds, 
if $r_R \sim r_S \sim 10 \times b$.
This effect as $\sim O(10^2)$ microarcseconds is 
marginally within the current VLBI accuracy. 
Therefore, the Weyl gravity model in the parameter region 
$\gamma \sim (r_H)^{-1}$ 
may be relevant with the current observation. 
However, Eddington-type measurements (which use the source and Earth motion 
and make a comparison between a lensed image position and an unlensed position) 
cannot be used for a light source such as a quasar at cosmological distance. 
Further investigations of how to test the $m\gamma$ coupling term 
in Weyl gravity by lensing observations at cosmological distance are left for future.

Interestingly, effects of finite distance in Weyl gravity 
may be relevant to the current observation, 
though those in Kottler are negligible as discussed in Section III.A. 
Why does such a crucial difference occur? 
The reason is the dependence on $r_H$. 
For the simplicity, we assume that $bU$ is small but not so negligible, 
for instance $bU \sim 1/10$. 
Eq. (\ref{V}) is roughly approximated as 
\begin{align}
V(U) \sim 
\frac16 
\left(\frac{r_g}{r_H}\right)^2 
\left(\frac{b}{r_g}\right) , 
\label{V-approx}
\end{align}
where $\Lambda \sim (r_H)^{-2}$ and we neglected $b^2U^2 \sim 10^{-2}$ . 
On the other hand, 
Eq. (\ref{W}) 
is roughly approximated as 
\begin{align}
W(U) \sim - \frac15 \left(\frac{m}{r_H}\right) 
\left(\frac{bU}{1/10}\right) , 
\label{W-approx}
\end{align}
where $\gamma \sim (r_H)^{-1}$. 
$V(U)$ is significantly suppressed by 
the inverse square of $r_H$ and thus 
it is negligible, 
while $W(U)$ is proportional to the inverse of $r_H$ and 
thus it is mildly small but not so negligible.

For readers' convenience, 
we rearrange Eq. (\ref{alpha-Weyl}) as 
\begin{align}
\alpha_{Weyl} = \alpha_{Weyl}^{\infty} + \alpha_{Weyl}^F , 
\label{alpha-Weyl2}
\end{align}
where $\alpha_{Weyl}^{\infty}$ is $\alpha_{Weyl}$ at infinity ($r_R, r_S \to \infty$) 
and $\alpha_{Weyl}^F$ denotes the finite-distance effect on the light deflection. 
\begin{align} 
\alpha_{Weyl}^{\infty} 
=&
  \frac{4m}{b} + O\left(m^2, \gamma^2\right) . 
\label{alpha-Weyl-infty}
\end{align}
\begin{align} 
\alpha_{Weyl}^F
=&
  \frac{2m}{b}\left(\sqrt{1 - b^2 u^2_S} + \sqrt{1 - b^2 u^2_R} - 2\right)
\notag\\
&   - m\gamma\left(\frac{bu_S}{\sqrt{1 - b^2 u^2_S}} 
   + \frac{bu_R}{\sqrt{1 - b^2 u^2_R}}\right)
  \notag\\
&  + O\left(m^2, \gamma^2\right) . 
\label{alpha-Weyl-F}
\end{align}

\section{Conclusion}
From the receiver's viewpoint, we made an attempt to provide 
a physical interpretation of the deflection angle defined by Ishihara et al. 
\cite{Ishihara2016}. 
This interpretation does not need the asymptotic flatness. 
Therefore, this interpretation encouraged us to seek 
another integral form of the deflection angle of light. 
The proposed integral form of the deflection angle can be used not only 
for asymptotically flat spacetimes but also for asymptotically nonflat ones. 
By doing explicit calculations, 
we examined the proposed deflection angle 
in two asymptotically nonflat spacetime models; the Kottler solution and 
a spherical solution in Weyl conformal gravity.
 
According to the present order-of-magnitude estimate, 
the extra deflection angle in Weyl gravity is within accuracy of the current VLBI, 
if some parameter values are in a certain range. 
Further investigations of how to test Weyl gravity by lensing observations 
are needed. 
On the other hand, effects of finite distance in Kottler spacetime can be safely ignored. 

Extensions to a case without spherical symmetry and so on are left for future.

\begin{acknowledgments}
We are grateful to Marcus Werner for the useful discussions. 
We wish to thank Kimet Jusufi for the helpful comments on his works. 
We would like to thank 
Yuuiti Sendouda, Ryuichi Takahashi, Kei Yamada and Masumi Kasai
for the useful conversations. 
We thank Ryunosuke Kotaki, Masashi Shinoda and Hideaki Suzuki 
for discussions. 
This work was supported 
in part by Japan Society for the Promotion of Science (JSPS) 
Grant-in-Aid for Scientific Research, 
No. 17K05431 (H.A.),  No. 18J14865 (T.O.), 
in part by Ministry of Education, Culture, Sports, Science, and Technology,  
No. 17H06359 (H.A.)
and 
in part by JSPS research fellowship for young researchers (T.O.).  
\end{acknowledgments}

\end{document}